\documentclass[preprint,pre,showpacs,floatfix]{revtex4}
\usepackage{graphicx}
\usepackage{dcolumn}
\usepackage{amsmath}
\usepackage{latexsym}
\usepackage{amssymb}
\usepackage{epsfig}
\usepackage{subfigure}

\newcommand{\stl}[1]{\mbox{$ \hspace{0.1em}
      \stackrel{\rule{0.4pt}{0.275ex}\hspace{0.40em} \!\!\!
      \overline{\hspace{0.06em}\vphantom{\rule{0.4pt}{0.0ex}}
      \hphantom{\mbox{$\displaystyle #1$}}
      \hspace{0.06em}  } \!\!\!\hspace{0.40em}\rule{0.4pt}{0.275ex}}
      {#1}\hspace{0.2em}$}}

\begin{document}

\title{Biaxiality  at the Isotropic-Nematic Interface with Planar Anchoring}

\author{S. M. Kamil, A. K. Bhattacharjee, R. Adhikari and Gautam I. Menon}
\affiliation{
The Institute of Mathematical Sciences, CIT Campus, 
Taramani, Chennai 600 113, India
}
\date{\today}
\begin{abstract}
We revisit the classic problem of the structure of the  isotropic-nematic interface within 
Ginzburg-Landau-de Gennes theory, refining previous analytic treatments of biaxiality 
at the interface. We compare our analysis with numerical results obtained through a
highly accurate spectral collocation scheme  for the
solution of the Landau-Ginzburg-de Gennes equations. In comparison to earlier work, we obtain 
improved agreement with numerics for both the uniaxial and biaxial profiles,  accurate asymptotic 
results for the decay of biaxial order on both nematic and isotropic sides of the interface
and  accurate fits to data from density functional approaches to this
problem.
\end{abstract} 
\maketitle

Liquid crystalline states of matter provide a useful testing ground 
for statistical mechanical theories of interface structure, since a variety of 
ordered phases can be accessed in experiments and computer simulations. 
The structure of the isotropic-nematic (I-N) interface presents a simple example of how interfacial order 
can differ radically from order in the coexisting bulk phases, since biaxial order is generically expected at
the interface even if the stable ordered phase is purely uniaxial. The study of 
the isotropic-nematic interface was initiated in an insightful paper by de Gennes, who 
introduced  a simple uniaxial ansatz for the tensor order parameter $Q_{\alpha\beta}$
which describes nematic order \cite{degennes}. The de Gennes ansatz is exact in the absence of elastic 
anisotropy. However, the description of the interface in the presence of such anisotropy 
poses a formidable analytic and numerical problem, since the partial differential equations for the 
five independent components  of  $Q_{\alpha\beta}$  contain non-linear couplings,  while $Q_{\alpha\beta} $ 
is itself constrained by symmetry and the requirement that its trace vanish.
 
Popa-Nita, Sluckin and Wheeler (PSW) \cite{popasluckwh} studied the I-N interface incorporating elastic anisotropy 
in the limit of planar anchoring, adapting a  parametrization introduced by Sen and Sullivan\cite{sensullivan}. In this parametrization,
the principal axes of $Q_{\alpha\beta}$ remain fixed in space,  and the problem reduces
to the solution of  two coupled non-linear partial differential equations in the dimension perpendicular
to the interface. These  equations represent  the variation of the amplitude of
uniaxial and biaxial ordering across the interface.  PSW showed that  the solutions of these equations
exhibited biaxiality in a region about the interface \cite{popasluckwh}. The uniaxial order parameter (S) was adequately 
represented by a tanh profile, as in the original calculation of de Gennes, while the biaxial order parameter (T)
exhibited more complex behaviour, peaking towards the isotropic side and with a trough on the 
nematic side. The biaxial profile was also shown to have a long tail towards the isotropic side, a feature 
hard to anticipate on physical grounds. 

This paper extends these calculations in several new ways. First, we show that
terms dropped by PSW in their simplification of the Ginzburg-Landau-de Gennes (GLdG)  equations
are,  in fact, comparable  in magnitude to the terms they retain, especially for small values of
$\kappa = L_2/L_1$, the ratio of the coefficients of the two lowest-order gradient terms in the GLdG expansion. Thus, a more accurate
treatment of the interface requires that these terms be retained.  The resulting equations have closed form 
solutions in terms of hypergeometric functions. We show  that such solutions provide a better description 
of the numerical data than the original calculation of PSW. We benchmark our analytic results through an accurate numerical
procedure, based on a Chebyshev polynomial expansion, for the study of these equations. 

We begin with the GLdG expansion of the free energy for a general $Q_{\alpha\beta}$
\begin{eqnarray}
{\mathcal F} &=& \int dz d{\bf x}_\perp  [ \frac{1}{2}ATr{\bf Q}^{2} + \frac{1}{3}BTr{\bf Q}^{3} +
\frac{1}{4}C(Tr{\bf Q}^{2})^{2} \nonumber \\
&&+ \frac{1}{2}L_1(\partial_\alpha Q_{\beta \gamma})(\partial_\alpha Q_{\beta \gamma}) + 
\frac{1}{2}L_2(\partial_\alpha Q_{\alpha \gamma})(\partial_\beta Q_{\beta \gamma})].
\label{freeg}
\end{eqnarray}
Here $A,B$ and $C$ are expansion parameters, while  $L_1, L_2$ are elastic constants. We choose
$B = -0.5, C= 2.67$ and $A = B^2/27C$, thus enforcing phase coexistence between an isotropic and
uniaxial nematic phase \cite{gralondej}. The interface is
taken to be flat and infinitely extended in the $x-y$ plane. 
The spatial variation of the order parameter only occurs along the $z$ direction\cite{sensullivan}. 
We scale $Q_{\alpha \beta} \rightarrow  Q_{\alpha \beta}/S_c$ where $S_c = -\frac{2B}{9C}$,
$\mathcal{F}  \rightarrow  \frac{16}{9CS_c^4}\mathcal{F}$, and measure lengths in units of
$l_c = \sqrt{54 C (L_1 + 2L_2/3)/B^2}$; we choose $L_1 = 10^{-6}$ in our numerics and obtain
$L_2$ from our choice of $\kappa$.
In the case of planar anchoring, the ordering at infinity is purely uniaxial and taken to be along the $x$ axis. In this
case, as shown by Sen and Sullivan,  uniaxial and biaxial order vary only with
$z$ and the principal axes of the {\bf Q} tensor remain fixed in space. The form of ${\bf Q}$ is then
\begin{equation}
{\bf Q} =\left ( \begin {matrix}S\ &0 &0 \\0&\frac {1} {2} (-S +  T) & 0\\0 & 0 &-\frac{1}{2}(S +  T)\end {matrix} \right).
\label{tpQ}
\end{equation}
\begin{figure}

\includegraphics[width=5in]{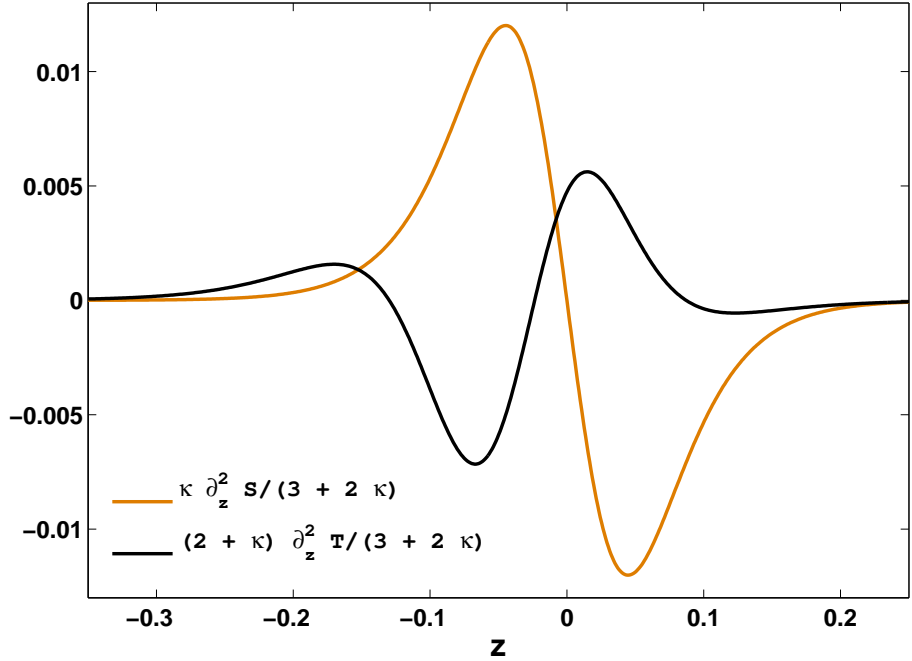}
\caption{(Color online) A comparison of the terms $\frac{(2 + \kappa)}{(3 + 2 \kappa)}\partial_z^2T$  (dark line) and
$\frac{\kappa}{(3 + 2 \kappa)}\partial_z^2 S$ (light line) obtained within the PSW solution to the GLdG equations,
for a $\kappa$ value of 18.0. The PSW approximation consists of ignoring the  $\frac{(2 + \kappa)}{(3 + 2 \kappa)}\partial_z^2T$ term
in comparison to the $\frac{\kappa}{(3 + 2 \kappa)}\partial_z^2 S$ term. Both terms, however, are of comparable magnitude.}
\label{dropped}
\end{figure}

Inserting this form of ${\bf Q}$ into the free energy and performing the minimization yields \cite{popasluckwh}
\begin{eqnarray}
\frac{(6 + \kappa)}{(3 + 2 \kappa)}\partial_z^2S +\frac{\kappa}{(3 + 2 \kappa)}\partial_z^2 T & = & 4 S - 12 S^2 + 8 S^3 + 4 T^2 + \frac{8ST^2}{3},
\label{ndfirst1}
\end{eqnarray}
\begin{eqnarray}
\frac{\kappa}{(3 + 2 \kappa)}\partial_z^2 S +\frac{(2 + \kappa)}{(3 + 2 \kappa)}\partial_z^2T & = & \frac{4}{3} T + 8 S T + \frac{8 T^3}{9} + \frac{8S^2 T}{3}.
\label{ndsecond2}
\end{eqnarray}

Popa-Nita, Sluckin and Wheeler now make several approximations to Eqs.~\ref{ndfirst1} and \ref{ndsecond2} to solve them. First, in Eq.~\ref{ndfirst1}, all terms in $T$ 
are dropped, since $S$ is typically much larger than $T$. The resulting equation for $S$ is solved by the tanh function. In Eq.~\ref{ndsecond2},
PSW drop the $\frac{(2 + \kappa)}{(3 + 2 \kappa)}\partial_z^2T$ term while retaining 
$\frac{\kappa}{(3 + 2 \kappa)}\partial_z^2 S$.  A test of self-consistency of this approximation is the comparison 
of the magnitude of these terms within the theory. Fig.~\ref{dropped} shows the terms $\frac{(2 + \kappa)}{(3 + 2 \kappa)}\partial_z^2T$  (dark line) 
and $\frac{\kappa}{(3 + 2 \kappa)}\partial_z^2 S$ (light line) computed through the PSW solution.
As can be seen  from the figure these terms only differ by a factor of order unity. 
Deep into the isotropic side, the term ignored by PSW exceeds the value of the term retained. 
Thus, while the PSW approach leads to a straightforward  algebraic relation
between $T$ and $S$, a more accurate method would be to retain the partial derivative term as well, requiring
that we solve a partial differential equation as opposed to an algebraic one. 

\begin{figure}
\subfigure{\label{fig:AKSTnew0_0}
\centering
\includegraphics[width=3.0in, height=2.6in, angle=0]{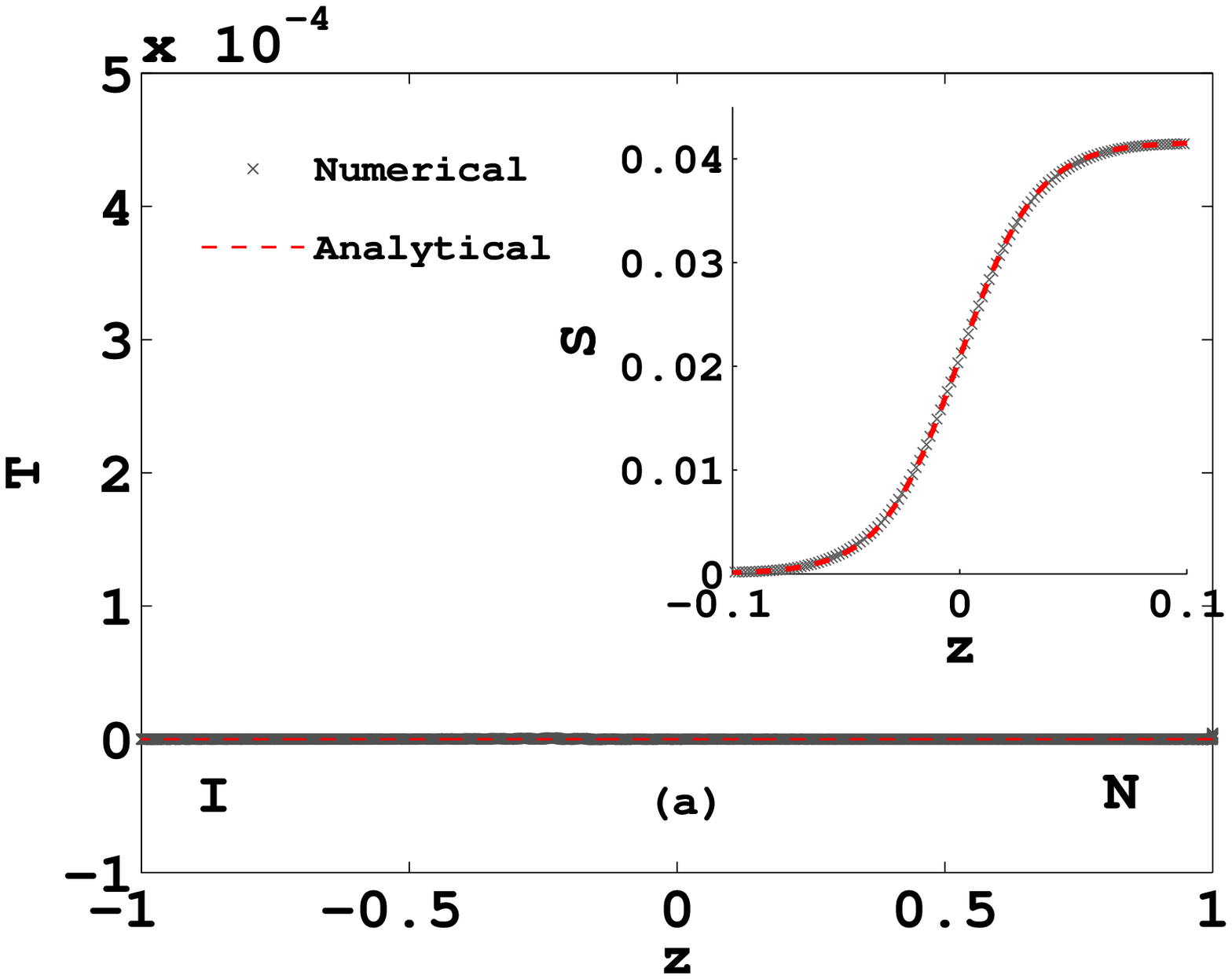}
}
\subfigure{\label{fig:AKSTnew0_4}
\centering
\includegraphics[width=3.0in, height=2.6in, angle=0]{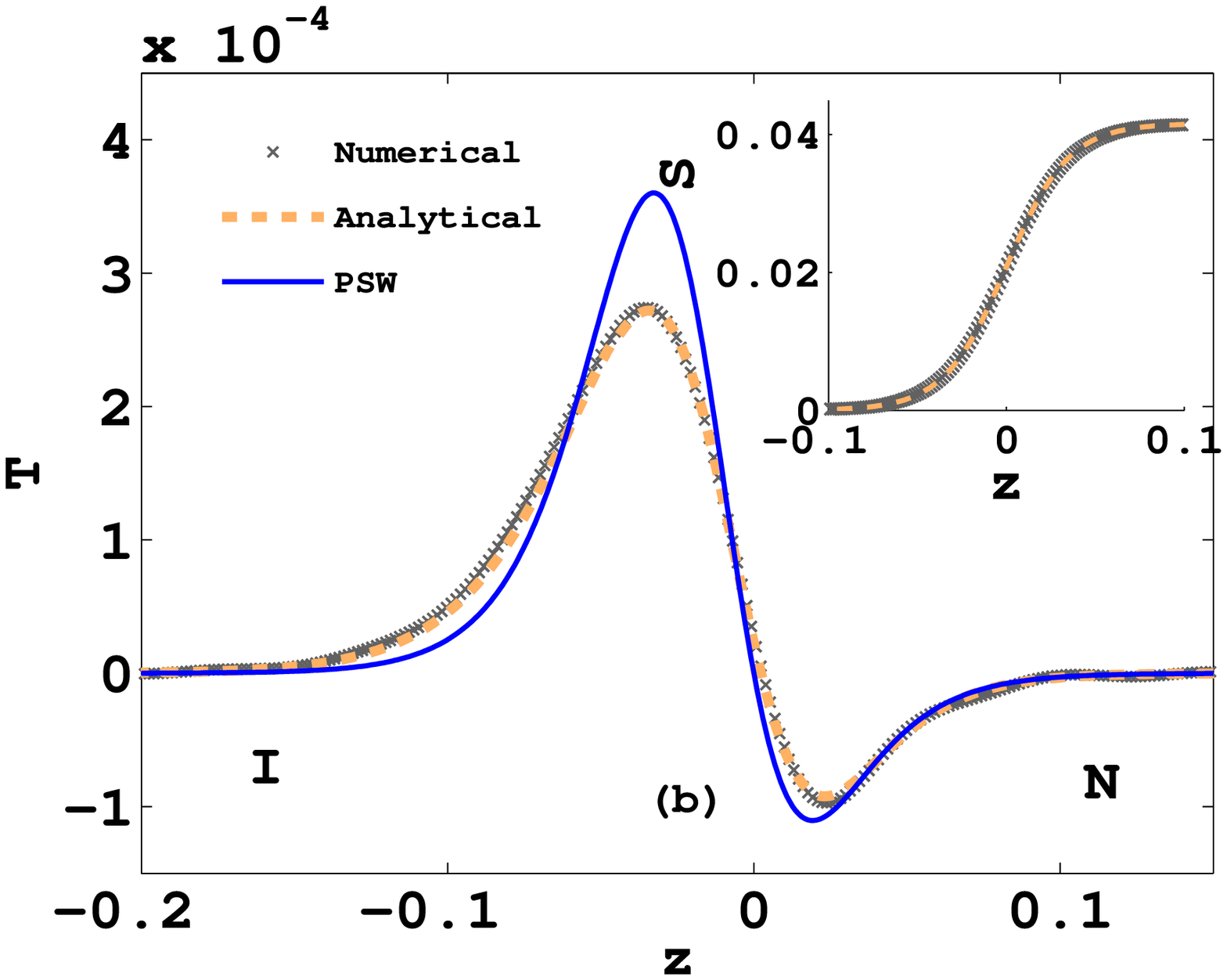}
}
\subfigure{\label{fig:AKSTnew4}
\centering
\includegraphics[width=3.0in, height=2.6in, angle=0]{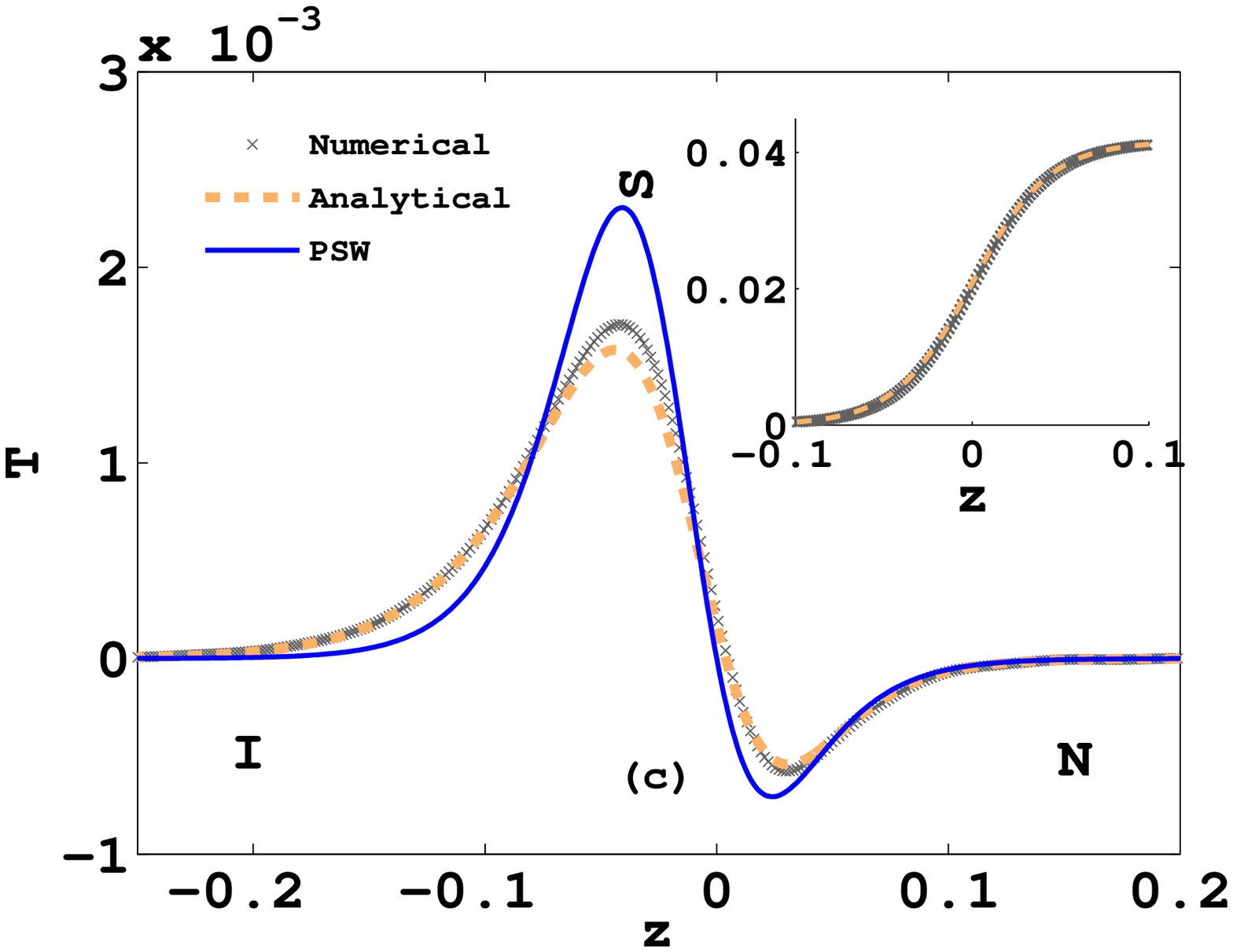}
}
\subfigure{\label{fig:AKSTnew18}
\centering
\includegraphics[width=3.0in, height=2.6in, angle=0]{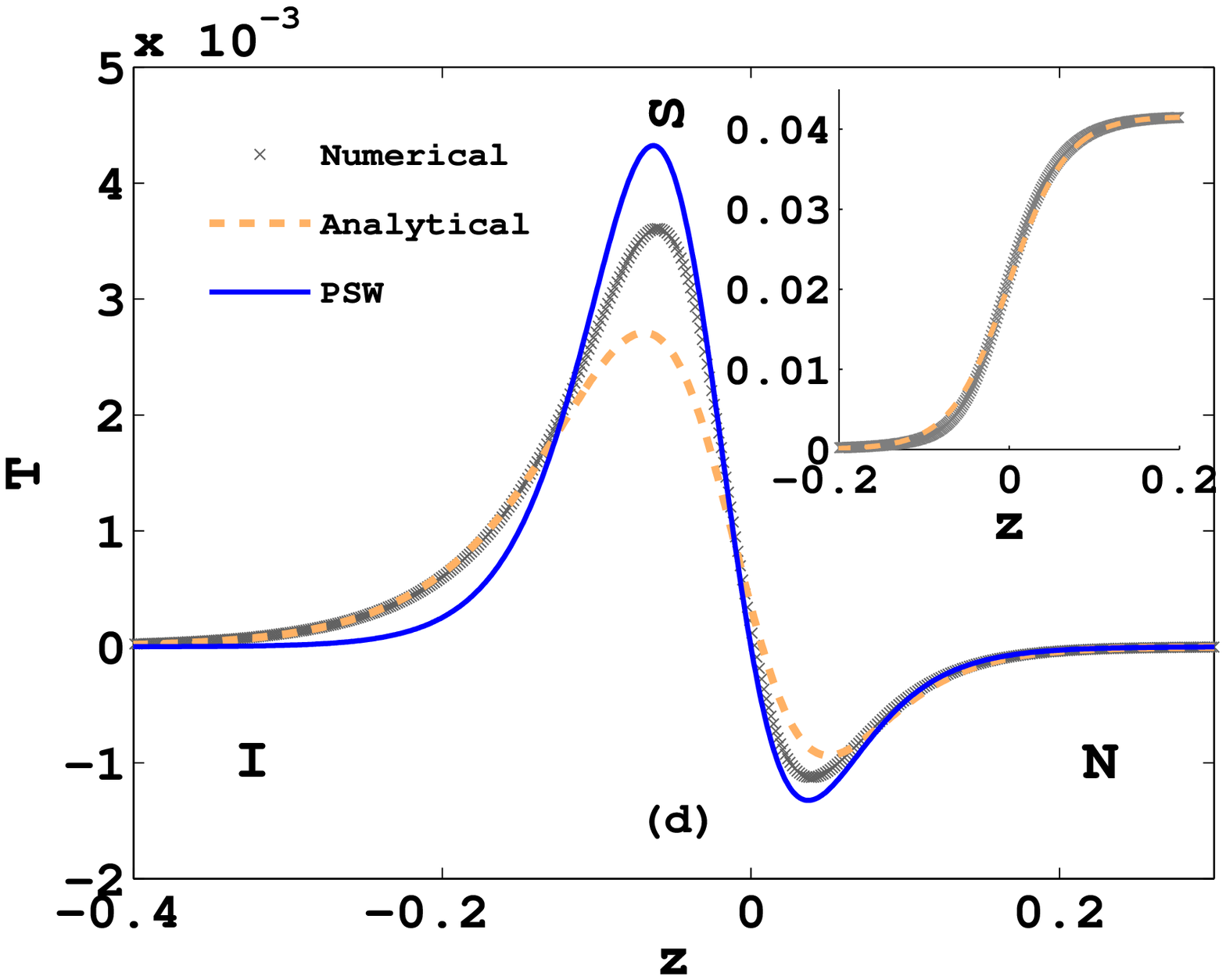}
}
\caption{(Color online) Biaxial and uniaxial profiles for $\kappa = 0 (a), 0.4 (b), 4 (c)$ and $18.0 (d)$, comparing results from our numerical 
computations ($\times$), with our analytic formula (dashed line)
and the formula of PSW (solid line).  The main figure shows the biaxial profile whereas the inset shows the uniaxial 
profile. In (a), for $\kappa = 0$, the solution has $T=0$, with the S profile exactly given
by the tanh form. In (b), for $\kappa = 0.4$,  the computed  biaxial profile (T) (main panel) is fit remarkably  well by our 
analytic form,  whereas  the PSW approximation tends to overestimate the peak value. The uniaxial (S) profile
 is shown in the inset of (a); here the results obtained by us and by PSW are
identical and the fit to a tanh profile is accurate over the entire region. In (c) (main panel), for $\kappa = 4.0$,  the numerical data are fit  
well by the analytic forms, particularly away from the main peak, yielding essentially exact agreement deep into the isotropic and 
nematic sides.  The PSW approximation is still an  overestimate to the peak value, and also differs sharply in  relation to the 
numerical data deep into the isotropic side.  The inset  shows the uniaxial (S) profile for this case. In (d) (main panel), for $\kappa =18.0$,
the PSW form appears to fit the peak better for larger $\kappa$, but again fails to capture the decay towards the isotropic side.}
\label{combined}
\end{figure}

\begin{figure}[h]
\begin{center}
\includegraphics[width=5in]{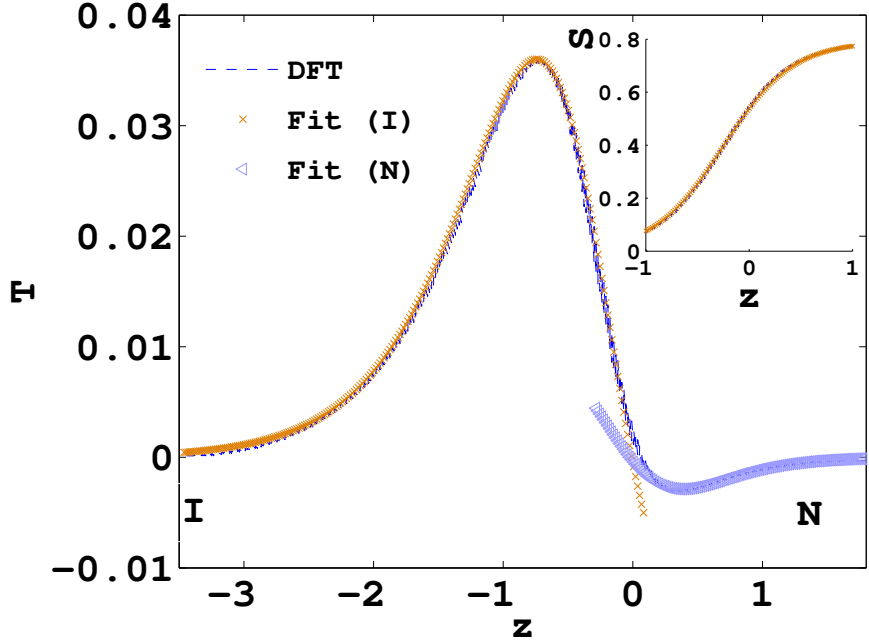}
\end{center}
\caption{(Color online) A comparison of  the results of our analytic calculation to profiles of $T$ obtained from a density
functional calculation for the isotropic-nematic interface. Profiles obtained for two values of $\kappa$,
$\kappa = 5.8$ (for $ z < 0$) and $\kappa = 0.69$ (for $z > 0$) are shown. The larger $\kappa$ value essentially
fits the $T$ profile exactly on the isotropic side, whereas the smaller $\kappa$ value provides an accurate fit
on the nematic side. The inset shows the $S$ profile obtained from the density functional calculation, together
with an optimum fit varying the value of $l_c$}
\label{dftcomp}
\end{figure}

Our approach to this problem uses the same approximations as PSW for Eq.~\ref{ndfirst1}. 
We thus take 
\begin{equation}
S = \frac{S_c}{2}\left [1 + \tanh(\frac{z}{\sqrt{2}\xi})\right ],
\end{equation}
where $\xi = \sqrt{\frac{1 +\kappa/6}{1+2\kappa/3}} $. Inserting this in equation (\ref{ndsecond2}),
scaling $z$  by $\sqrt{2} \xi$, redefining the resulting quantity as $z$ again, and  dropping  the nonlinear term, we obtain,
\begin{equation}
\partial^2_z T  = 2\beta [\tanh^2(z) + 8 \tanh(z) + 9 ]T 
+ \frac{\kappa}{2+\kappa}\tanh(z)[1+ \tanh(z)][1 - \tanh(z)].
\end{equation}
with $\beta = \frac{6+\kappa}{3(2+\kappa)}$.

The solution of the equation consists of a homogeneous part $T_h$ and a particular part $T_p$
where
$T_h  = C_1 t^{3\sqrt{\beta}}(1-t)^{-\sqrt{\beta}}{}_2F_1[a_1,b_1,c_1,t] + C_2 t^{-3\sqrt{\beta}}(1-t)^{\sqrt{\beta}}{}_2F_1[a_2,b_2,c_2,1-t]$
and 
$a_1 	=  \frac{1}{2} + 2\sqrt{\beta} + \frac{\sqrt{1 + 8\beta }}{2};\quad b_1 = \frac{1}{2} + 2\sqrt{\beta} - \frac{\sqrt{1 + 8\beta }}{2};\quad c_1 = 1+ 6\sqrt{\beta};
a_2 	=  \frac{1}{2} - 2\sqrt{\beta} - \frac{\sqrt{1 + 8\beta }}{2};\quad b_2 = \frac{1}{2} - 2\sqrt{\beta} + \frac{\sqrt{1 + 8\beta }}{2};\quad c_2 = 1+ 2\sqrt{\beta}$
and  $ t =\Big(\frac{1 - \tanh(z)}{2}\Big)$. The function  ${}_2F_1 $ is a hypergeometric function and  $C_1$ and $C_2$ are fixed by boundary
conditions. 

The particular solution takes the form
\begin{equation}
T_p(z) = \left [ -y_1(z) I_2(z) + y_2(z) I_1(z) \right ]/W(z),
\label{maineq}
\end{equation}
with
\begin{eqnarray}
y_1(z) & = &\Big(\frac{1 - \tanh(z)}{2}\Big)^{3\sqrt{\beta}}\Big(\frac{1+\tanh(z)}{2}\Big)^{-\sqrt{\beta}}{}_2F_1[a_1,b_1,c_1,\frac{1 - \tanh(z)}{2}]\nonumber \\
y_2(z) & = &\Big(\frac{1-\tanh(z)}{2}\Big)^{-3\sqrt{\beta}}\Big(\frac{1+\tanh(z)}{2}\Big)^{\sqrt{\beta}}{}_2F_1[a_2,b_2,c_2,\frac{1 + \tanh(z)}{2}]
\label{homog}
\end{eqnarray}
and the Wronskian $W(z) = W = y_1 (d y_2/dz) - y_2 (dy_1/dz)$, where
\begin{eqnarray}
I_1(z) & = & \frac{2 \kappa}{2+\kappa} \sum_{m=0}^{\infty}\frac{(a_1)_m(b_1)_m}{(c_1)_m m!} t^{1+m +3\sqrt{\beta}}(1-t)^{1-\sqrt{\beta}}\Big(-\frac{2}{2+m+2\sqrt{\beta}}\nonumber \\
& & + \frac{m+ 4\sqrt{\beta}}{(2+m+2\sqrt{\beta})(1+m+3\sqrt{\beta})}{}_2F_1[1 ,2+m+2\sqrt{\beta},2+m+3\sqrt{\beta},t]\Big)\nonumber \\
\end{eqnarray}
\begin{eqnarray}
I_2(z) & = & \frac{2 \kappa}{2+\kappa} \sum_{n=0}^{\infty}\frac{(a_2)_n(b_2)_n}{(c_2)_n n!} t_1^{1+n +\sqrt{\beta}}(1-t_1)^{1-3\sqrt{\beta}}\Big(-\frac{2}{2+n-2\sqrt{\beta}}\nonumber \\
& & + \frac{n+ 4\sqrt{\beta}}{(2+n-2\sqrt{\beta})(1+n+\sqrt{\beta})}{}_2F_1[1 ,2+n-2\sqrt{\beta},2+n+\sqrt{\beta},t_1]\Big).\nonumber \\
\end{eqnarray}
The Pochhammer symbol $(a)_n$ which enters above is defined via $(a)_n = a(a+1)(a+2)\ldots (a+n-1)$. 
Here $t_1 =  [1 + \tanh(z)]/2$ and the result for $I_1(z)$ and $I_2(z)$ is obtained by expanding the hypergeometric functions in
Eqns.~\ref{homog} in a power series and integrating term-by-term\cite{abramstegun}.  Note that
the solutions of the homogeneous part diverge asymptotically. Thus,  for the
boundary condition $T = 0$  at $z = \pm\infty$
the only physical solution is the particular one. Eq.~\ref{maineq} is thus the key analytical result of this paper, describing the
variation of  biaxiality across the interface. In our numerical evaluations, we sum the series for $I_1(z)$ and $I_2(z)$, retaining
as many terms as are required to ensure convergence. 
The  series in $I_2$ converges very fast (only 3 terms need be retained for good results) whereas the  series in $I_1$ converges more slowly
and around 9 terms must be retained for convergence.
To convert these into physical units, we must undo the sequence of length transformations, replacing 
$z \rightarrow z /(\sqrt{2} \xi l_c)$.

An asymptotic analysis of these equations is possible: for $z \rightarrow -\infty$, $S$ and $T$ are  small.
The  $\tanh$ profile for $S$ can be approximated as
$\frac{1}{2}(1+\tanh(\frac{z}{\sqrt{2}\xi})) \rightarrow e^{\frac{2z}{\sqrt{2}{\xi}}}$
while Eq.~(\ref{ndsecond2}) takes the form
$2 \xi^2 \partial^2_z T  =  4\beta T - (\frac{2\kappa}{2+\kappa}) e^{2 \frac{z}{\sqrt{2}\xi}}$
with $\beta = \frac{6+\kappa}{3(2+\kappa)}$.
Thus
$\partial^2_z T  =  \frac{4}{3}\frac{(3+2\kappa)}{(2+\kappa)} T -\frac{2\kappa(3+2\kappa)}{(2+\kappa)(6+\kappa)} e^{2 \frac{z}{\sqrt{2}\xi}}$
with  asymptotic solution
\begin{equation}
T \sim e^{\sqrt{\frac{4(3+ 2\kappa)}{3(2+\kappa)}}z}, z \rightarrow -\infty,
\end{equation}
a result in perfect accord with the computed
forms of $T$ deep into the isotropic phase. Note that  $T > S$ and $ (S+T)/2 > S$ as one moves deeper into the  isotropic side. 
This implies that the principal order parameter is negative as pointed out in  Ref.~\cite{popasluckwh}, where this result was 
obtained numerically.

As $ z\rightarrow \infty$, an alternative asymptotic expansion can be derived by taking
$S = 1 - \frac{1}{2} e^{-bx}$
with $b = 2\sqrt{\frac{3 + 2 \kappa}{6 + \kappa}}$.
We then obtain
\begin{equation}
T \sim e^{- 2\sqrt{\frac{3 + 2 \kappa}{6 + \kappa}} z},  z \rightarrow \infty,
\end{equation}
in agreement with our numerical  results. Popa-Nita, Sluckin and Wheeler provide an  analysis of the 
asymptotics in the specific limit  that $\kappa \rightarrow \infty$. However, our results cannot be directly 
translated to this limit, since we assume a tanh profile of S; this approximation becomes increasingly
inaccurate for larger $\kappa$ (see below). 
 
Our numerical results are obtained using a spectral collocation method \cite{trefethen}, applied to our knowledge 
for the first time to the GLdG equations. In the spectral collocation, the solution is expanded in an orthogonal 
basis of Chebyshev polynomials in a bounded interval. Differentiation operators  constructed from this Chebyshev 
interpolant are spectrally accurate, in the sense that the error vanishes exponentially in the number of retained 
polynomials. The interpolant is constructed so as  to satisfy  Dirichlet boundary conditions. Though 
the physical problem is for an unbounded interval, our numerical approximation of a bounded interval  gives excellent 
results since all  variation in the order parameters is restricted to the region proximate to the interface. 

Specifically, we solve  the equations of equilibrium
\begin{equation}
(A + C TrQ^{2})Q_{\alpha\beta}({\bf x}, t) + B
\stl{Q^{2}_{\alpha\beta}({\bf x}, t)} = L_{1}\nabla^{2}Q_{\alpha\beta}({\bf x}, t)
 + L_{2}\stl{\nabla_{\alpha}(\nabla_{\gamma}Q_{\beta\gamma}({\bf x}, t))}
\end{equation}
by  transforming to a basis $\{a_i\}$ which enforces symmetry and tracelessness, as  $Q_{\alpha\beta} = 
\sum_{i=1}^{5}a_{i}T^{i}_{\alpha\beta}$, where, ${\bf T}^{1} = \sqrt{\frac{3}{2}} \stl{{\bf \hat{z}\hat{z}}},
{\bf T}^{2} =  \sqrt{\frac{1}{2}} ({\bf \hat{x} \; \hat{x} - \hat{y} \; \hat{y}}),{\bf T}^{3} = \sqrt{2}\; 
\stl{{\bf \hat{x} \; \hat{y}}},{\bf T}^{4} = \sqrt{2}\; \stl{{\bf \hat{x} \; \hat{z}}}$, ${\bf T}^{5} = 
\sqrt{2}\; \stl{{\bf \hat{y} \; \hat{z}}}$. Overbars indicate traceless symmetric parts. We thus obtain five simultaneous partial differential equations 
for the $a_i$, which are steady-states of the time-dependent equations we have obtained earlier \cite{bhmead}. Note specifically
that we make no symmetry-based ansatz for the components of $Q_{\alpha \beta}$ \cite{footnote}. 

The spectral collocation
reduces these differential equations to non-linear algebraic equations. We solve them 
using a relaxation method from a well-chosen initial condition, relaxing till the differential change in successive 
iterations is less than $10^{-5}$. Spectral convergence to machine accuracy is obtained by retaining 
128 Chebyshev modes, as we have checked by an explicit calculation. 
To compare with analytical and density functional results, the solution at the Chebyshev nodes is interpolated 
using barycentric interpolation without compromising spectral accuracy. The DMSUITE library is used for the numerical 
implementation \cite{weired}.

Our results are summarized in Fig.~\ref{combined} and Fig.~\ref{dftcomp}.  The main panel of
Fig.~\ref{combined} (a), obtained by solving  Eq.~\ref{ndfirst1} and \ref{ndsecond2} for a
value of $\kappa = 0.0$, shows the biaxiality profile obtained using our numerical spectral scheme (crosses), as compared to
the analytic result of $T=0$. The uniaxiality profile shown in the inset is exactly the tanh profile obtained by
de Gennes.  This limit provides a simple test of our numerical methods, since the solution  to Eqs.~\ref{ndfirst1} and
\ref{ndsecond2} in this limit is exact. Fig.~\ref{combined} (b) shows the biaxiality profile obtained using our spectral scheme (crosses), as compared to
the analytic results derived here (dashed line) and results obtained by PSW (solid line) for a value of 
$\kappa = 0.4$. As can be seen, the
numerical data are fit remarkably well by the analytic forms, whereas the PSW approximation tends to overestimate
the peak value. The inset to Fig.~\ref{combined} (b) shows the uniaxial (S) profile, obtained numerically as well as in
our analytic calculation; here the results obtained by us and by PSW are identical. The fit to a tanh profile is
accurate over the entire region. 

The main panel of Fig.~\ref{combined} (c) shows the biaxiality profile obtained using our spectral scheme (crosses), as compared to
the analytic results derived here (dashed line) and results obtained by PSW (solid line) for a value of
$\kappa = 4$. Again the
numerical data are fit  well by the analytic forms, particularly away from the main peak, yielding essentially
exact agreement deep into the isotropic and nematic sides.  The PSW approximation is still an overestimate
to the peak value, and also differs sharply in  relation to the numerical data deep into the isotropic side. 
The inset to Fig.~\ref{combined} (c) shows the uniaxial (S) profile for this case.
Fig.~\ref{combined} (d) shows the biaxiality profile obtained using our spectral scheme (crosses), as compared to
the analytic results derived here (dashed line) and results obtained by PSW (solid line) for a value of
$\kappa = 18$. For these - and larger - values of $\kappa$, our analytic fits differ noticeably from the
numerical data. The PSW form appears to fit better for larger $\kappa$, although we believe that this
is fortuitous. It appears that the principal error arises from our approximation of the $S$ profile as
a tanh form. For large $\kappa$, this approximation is less accurate. 

Fig.~\ref{dftcomp} compares the results of our analytic calculation to profiles of $T$ obtained from a density
functional calculation for the isotropic-nematic interface \cite{chen2} a method which  provides an alternative, more
molecular approach to this problem\cite{schilling}. We have taken numerical data for uniaxial and 
biaxial profiles obtained in Ref.~\cite{chen2}, varying the free parameters $S_c$, $l_c$ and $\kappa$ in our
solutions till an optimal fit is obtained. The values of $S_c$ and $l_c$ can be obtained from fits to $S$; thus
only $\kappa$ need be varied to represent the $T$ profile.  Fig. ~\ref{dftcomp} shows profiles obtained for two values of $\kappa$:
$\kappa = 5.8$ (for $ z < 0$) and $\kappa = 0.69$ (for $z > 0$). The larger $\kappa$ value 
fits the profile very closely on the isotropic side, whereas the smaller $\kappa$ value provides an accurate fit
on the nematic side \cite{footnote1}. It does not seem possible to fit the complete profile using a single value of $\kappa$. 
This  could have been anticipated on physical grounds since the density functional theory yields
a density difference between coexisting isotropic and nematic phases.  The elastic coefficients $L_1$ and $L_2$ which 
enter our calculation do in principle contain  a density dependence which we ignore here.

In conclusion, we have presented results for the uniaxial and biaxial profiles, in the case of
planar anchoring, for the  classic problem of the structure of the  isotropic-nematic interface within 
Ginzburg-Landau-de Gennes theory. Our work refines previous analytic treatments of biaxiality 
at the interface. We have implemented a  highly accurate spectral collocation scheme for the
solution of the Landau-Ginzburg-de Gennes equations and used this numerical scheme in our tests of
the analytic results. 

In comparison to earlier work, we obtain 
improved agreement with numerics for both the uniaxial and biaxial profiles, with our results being increasingly
accurate as the anisotropy  is reduced. We also provide accurate asymptotic results for the decay of the
$S$ and $T$ order parameters deep into the nematic and isotropic phases. Our calculated profiles show a pleasing 
consistency with profiles obtained from density functional approaches. Further extensions of these
numerical and analytic methods to the case of an intermediate anchoring condition far from the interface
are currently under way.

The authors thank C. Dasgupta for discussions. This research was supported in part by the DST(India)
and the Indo-French Centre for the Promotion of Advanced Research.

%%%%%%%%%%%%%%%%%%%%%%%


\begin{thebibliography}{13}
\bibitem[{foo({\natexlab{a}})}]{footnote}
\bibinfo{note}{The ambiguity of the sign of $T$ in the uniaxial phase, or
  whenever the ordering of the sub-leading eigenvalues changes, is resolved by
  requiring solutions to be smooth over the computational domain.}
\bibitem[{foo({\natexlab{b}})}]{footnote1}
\bibinfo{note}{An alternative density functional approach
  (Ref.~\cite{shundyak1}) yield profiles which can also be fit very well on the
  nematic side by our methods, although the fit towards the isotropic side is
  of reduced quality.}
\end{thebibliography}
\end{document}